\documentclass[aps,pra,twocolumn,superscriptaddress]{revtex4}
\usepackage{graphicx}
\usepackage{amsmath,amssymb}
\usepackage{bm}
\usepackage{subfigure}
\usepackage{color}
\usepackage{bbold}
\begin{document}
\title{Effects of dissipation on the superfluid-Mott-insulator transition of photons}
\author{A.-W. de Leeuw}
\email{A.deLeeuw1@uu.nl}
\affiliation{Institute for Theoretical Physics and Center for Extreme Matter and Emergent Phenomena, Utrecht
University, Leuvenlaan 4, 3584 CE Utrecht, The Netherlands}
\author{O. Onishchenko}
\affiliation{Institute for Theoretical Physics and Center for Extreme Matter and Emergent Phenomena, Utrecht
University, Leuvenlaan 4, 3584 CE Utrecht, The Netherlands}
\affiliation{Faculty of Applied Sciences, Delft University of Technology, Lorentzweg 1, 2628 CJ Delft, The Netherlands}
\author{R.A. Duine}
\affiliation{Institute for Theoretical Physics and Center for Extreme Matter and Emergent Phenomena, Utrecht
University, Leuvenlaan 4, 3584 CE Utrecht, The Netherlands}
\author{H.T.C. Stoof}
\affiliation{Institute for Theoretical Physics and Center for Extreme Matter and Emergent Phenomena, Utrecht
University, Leuvenlaan 4, 3584 CE Utrecht, The Netherlands}
\date{}
\date{\today}
\begin{abstract}
We investigate the superfluid-Mott-insulator transition of a two-dimensional photon gas in a dye-filled optical microcavity and in the presence of a periodic potential. We show that in the random-phase approximation the effects of the dye molecules, which generally lead to dissipation in the photonic system, can be captured by two dimensionless parameters that only depend on dye-specific properties. Within the mean-field approximation, we demonstrate that one of these parameters decreases the size of the Mott lobes in the phase diagram. By considering also Gaussian fluctuations, we show that the coupling with the dye molecules results in a finite lifetime of the quasiparticle and quasihole excitations in the Mott lobes. Moreover, we show that there are number fluctuations in the Mott lobes even at zero temperature and therefore that the true Mott-insulating state never exists if the interactions with the dye are included.
\end{abstract}
\maketitle
\vskip2pc
\section{Introduction}
\label{sec:int}
In physics there are many theoretically predicted phenomena that are hard to verify directly in experiments. This can have several reasons, such as that predictions are outside the limits of current devices or that other physics overshadows the desired effect. In the latter category examples are the effects of disorder on top of effects predicted for clean systems, or a  combination of various kinds of interactions. Therefore, there is a constant search for systems that exhibit interesting physics, are relatively simple, and are described by few parameters of which many are controllable experimentally.
\newline
\indent A prime example of such a system is obtained by combining cold atoms with an optical lattice. In this case there is almost full control over the interactions between the atoms and over the lattice structure. Therefore, there exists a broad variety of experimental possibilities in these systems that demonstrate many phenomena in condensed-matter physics, e.g. see Refs.\,\cite{optical1,optical2,optical3,optical4,optical5,optical6,optical7,optical8,optical9}. Nowadays this research area is still very active. One of the reasons is that cold fermionic atoms in an
optical lattice possibly are quantum simulators for high-temperature superconductivity \cite{htc1,htc2,htc3}.
\newline
\indent An interesting property of cold bosonic atoms in optical lattices is that there is a so-called quantum phase transition \cite{Sachdev}. By reducing the depth of the optical lattice the system undergoes a transition from the Mott insulator, where each site is filled with an equal and integer amount of particles and number fluctuations are suppressed, to the superfluid phase with a fluctuating number of atoms per site. This transition has first been observed experimentally in 2002 by Greiner \emph{et al.} \cite{smi1}, and hereafter it has been studied extensively both theoretically and experimentally \cite{smi2,smi3,smi4,smi5,smi6,smi7,smi8,smi9,smi10}. Although most studies focus on the superfluid-Mott-insulator transition for cold atoms, this phase transition is not restricted only to cold atomic gases. For example, it also has been studied in systems consisting of polaritons \cite{polar1,polar2,polar3,polar4,polar5,polar6,polar7}. Furthermore, the transition has been investigated in interacting photon gases in coupled dissipative cavities \cite{phot1,phot2,phot3,phot4}. 
\newline
\indent More recently, a new candidate for a system that can display a superfluid-Mott-insulator transition has emerged, namely photons in a dye-filled optical microcavity. After the observation of Bose-Einstein condensation of photons in this system \cite{BECphoton}, recently a new experiment was proposed \cite{PhotLat}. By periodically varying the index of refraction of the dye inside the cavity, an effective lattice potential for the photons can be induced. Therefore it is expected that the photons can also undergo this superfluid-Mott-insulator phase transition. However, this system is fundamentally different from the standard cold atomic gases in optical lattices, since the photons can be absorped and emitted by the dye molecules. Thus, the question arises how this coupling affects the behaviour of the photons in this periodic potential, and in particular the properties of the quantum phase transition to the Mott insulator.
\newline
\indent In this article we study the dissipation effects for a photonic lattice in a dye-filled optical microcavity. First, in Sec.\,\ref{sec:mod} we introduce the general theory and express all quantities that enter our theory in terms of experimentally known parameters. Subsequently, we determine the effect of the molecules on the Mott lobes in mean-field theory in Sec.\,\ref{sec:simmod}. We start by considering a simplified model that neglects the fixed longitudional momentum of the photons and only considers absorption and emission of photons with zero momentum. We show that at the mean-field level the dye effects are captured in one parameter $\gamma$ that can be calculated analytically, and we express this parameter in experimentally known quantities. Moreover, we show that incorporation of $\gamma$ decreases the size of the Mott lobes. Hereafter, we consider the model that includes the fixed longitudional momentum of the photons and also the absorption and emission of photons with an arbitrary momentum, and we study the effect of these extensions on the value of $\gamma$. In Sec.\,\ref{sec:numfluc} we go beyond mean-field theory and we calculate in the random-phase-approximation the excitations inside the Mott lobes. We show that in this approximation the dimensionless damping parameter $\alpha_{\mathrm{lat}}$ enters our model and therefore the excitations acquire a finite lifetime. Moreover, we show that even at zero temperature there are now number fluctuations inside the Mott lobes, which implies that the true Mott-insulating phase no longer exists if the interactions with the dye is taken into account. Finally, we end with a conclusion and outlook in Sec.\,\ref{sec:concl}.

\section{Photonic lattice in dye-filled microcavity}
\label{sec:mod}
In this section we write down a model for a lattice of photons in a dye-filled optical microcavity for the experimental set-up used in Ref.\,\cite{BECphoton,Nyman}. In the particular experimental configuration the longitudional momentum of the photons is fixed, and the photons behave equivalently to a massive harmonically trapped Bose gas in two dimensions. Since the photons interact with dye molecules, the imaginary-time action that describes the photon system contains three parts. First, the part that describes the photons reads
\begin{align}\label{eq:photact}
S&_{\mathrm{ph}}[\phi^{*}, \phi] = \int_{0}^{\hbar\beta} d\tau \int d{\bf x} \, \phi^{*}({\bf x}, \tau) \Bigg{\{}\hbar \frac{\partial}{\partial \tau} - \frac{\hbar^{2} \nabla^{2}}{2 m}   \\ \nonumber
&+ V^{\mathrm{ext}}({\bf x}) - \mu +  \frac{1}{2} \phi^{*}({\bf x}^{\prime}, \tau) V({\bf x} - {\bf x}^{\prime}) \phi({\bf x}^{\prime}, \tau) \Bigg{\}} \phi({\bf x}, \tau),
\end{align}
where $\mu$ is the chemical potential of the photons, $m$ denotes their effective mass, $\beta = 1/k_{\mathrm{B}} T$ with $T$ the temperature, $V({\bf x} - {\bf x}^{\prime})$ is the interaction potential and 
\begin{align}\label{eq:pot}
V^{\mathrm{ext}}({\bf x}) = V_{0} \sum_{j} \cos^{2}(2 \pi x_{j} / \lambda),
\end{align}
denotes the lattice potential with $\lambda$ two times the lattice spacing. In the following we are primarly interested how these photons are affected by the coupling to the molecules, and therefore we ignored the external harmonic potential that arises due to the curvature of the cavity mirrors. We model the dye as a two-level system with energy difference $\Delta$ and we introduce the effective mass $m_{\mathrm{d}}$ to model the rovibrational structure of the molecules. As is shown in Ref.\,\cite{AW}, the value of this effective mass can be tuned such that the correct experimental results for the molecular absorption and emission spectra are obtained. A different way to achieve this has been put forward by Kirton and Keeling \cite{Keeling}. Hence, the molecular part of the action reads
\begin{align}
S&_{\mathrm{mol}}[\psi^{*}, \psi] = \sum_{\rho \in\{\uparrow, \downarrow\}} \int_{0}^{\hbar\beta} d\tau \int d{\bf y} \, \psi_{\rho}^{*}({\bf y}, \tau) \\ \nonumber
&\times \Bigg{\{}\hbar \frac{\partial}{\partial \tau} - \frac{\hbar^{2} \nabla^{2}}{2 m_{\mathrm{d}}} + K_{\rho} - \mu_{\rho} \Bigg{\}} \psi_{\rho}({\bf y}, \tau),
\end{align}
where $K_{\uparrow} = \Delta$, $K_{\downarrow} = 0$ and $\mu_{\rho}$ denotes the chemical potential of the excited and ground-state molecules. Contrary to the photon part, here the integration is over three-dimensional space. From now on, we use the convention that ${\bf x}$ is a two-dimensional vector and ${\bf y}$ is three dimensional. The last part of the action consist of interaction terms between photons and molecules, and reads
\begin{align}
S_{\mathrm{c}}&[\psi^{*}, \psi,\phi^{*}, \phi] = g \int_{0}^{\hbar\beta} d\tau \int d{\bf y} \, \Big{\{}\phi^{*}({\bf y}, \tau) \\ \nonumber
&\times \psi_{\downarrow}^{*}({\bf y}, \tau) \psi_{\uparrow}({\bf y}, \tau) +  \phi({\bf y}, \tau) \psi^{*}_{\uparrow}({\bf y}, \tau) \psi_{\downarrow}({\bf y}, \tau)  \Big{\}} ,
\end{align}
with $g$ a coupling constant. Furthermore, $\phi^{*}({\bf y}, \tau)$ is related to the photon field $\phi^{*}({\bf x}, \tau)$ in Eq.\,\eqref{eq:photact} according to
\begin{align}
\phi({\bf y}, \tau) = \sqrt{2/ L} \sin\left(k_{\gamma} z \right) \phi({\bf x}, \tau).
\end{align}
Here we assumed that in one direction the photons are confined by a box of length $L$ with impenetrable barriers at either end. Furthermore, in agreement with the experiments we only take into account a single longitudional momentum $k_{\gamma}$. 
\newline
\indent To make further progress, we expand the photonic fields in Wannier functions as 
\begin{align}
\phi({\bf x}, \tau) = \sum_{{\bf n},i} a_{{\bf n},i}(\tau) w_{{\bf n}}({\bf x} - {\bf x}_{i}),
\end{align}
where $a_{{\bf n},i}(\tau)$ and its complex conjugate respectively annihilates or creates a photon in a Wannier state $w_{{\bf n}}({\bf x} - {\bf x}_{i})$ at lattice site $i$ and in the band with the band index ${\bf n}$. Furthermore, the molecular field is expanded as
\begin{align}
\psi_{\rho}({\bf y}, \tau) = \sum_{{\bf p}} b_{{\bf p},\rho}(\tau) \frac{e^{i {\bf p} \cdot {\bf y}}}{\sqrt{V}},
\end{align}
where $b_{{\bf p},\rho}(\tau)$ and $b^{*}_{{\bf p},\rho}(\tau)$ annihilates or creates a molecule with momentum ${\bf p}$ and internal state $|\rho\rangle$. Next we consider the tight-binding limit, where each site can be seen as a harmonic oscillator and the Wannier functions are known exactly. Furthermore, we consider the limit where the photons only occupy the lowest band ${\bf n} = {\bf 0}$. In this approximation the three parts of the action can be simplified, and by using the result of Ref.\,\cite{Dries} we obtain
\begin{align}\label{eq:photact}
&S_{\mathrm{ph}}[a^{*}, a] = - \int_{0}^{\hbar\beta} d\tau \sum_{i \neq j} a^{*}_{i}(\tau) t_{i,j} a_{j}(\tau) \\ \nonumber
&+ \int_{0}^{\hbar\beta} d\tau \sum_{i} a^{*}_{i}(\tau) \left\{\hbar \frac{\partial}{\partial \tau} + \epsilon_{i}  - \mu + \frac{U}{2} |a_{i}(\tau)|^{2} \right\} a_{i}(\tau),
\end{align}
where $\epsilon_{i}$ is the energy at lattice site $i$, $t_{i,j}$ is the hopping strength between sites $i$ and $j$ and $U$ is the on-site interaction strength. The expression of these quantities in terms of Wannier functions can be found in Ref.\,\cite{Dries}. Since we consider the tight-binding limit we know the analytic expression for $w_{{\bf 0}}({\bf x} - {\bf x}_{i})$. It is given by
\begin{align}\label{eq:harmos}
w_{{\bf 0}}({\bf x} - {\bf x}_{i}) = \left(\frac{m \omega}{\pi \hbar}\right)^{1/2} \exp\left\{- m \omega ({\bf x} - {\bf x}_{i})^{2} / 2 \hbar \right\},
\end{align}
where $m$ is the effective mass of the photon and $\omega$ is the frequency of the harmonic potential at every site that can be obtained by performing a Taylor expansion of Eq.\,\eqref{eq:pot}. Hence,
\begin{align}
\omega = \frac{2 \pi}{\lambda} \sqrt{\frac{2 V_{0}}{m}}.
\end{align}
Note that the single-band approximation is only valid if we have that both the thermal energy $k_{\mathrm{B}} T$ and the on-site interaction strength $U$ is smaller than the on-site energy of the photons $\hbar\omega$, i.e. $k_{\mathrm{B}} T \ll \hbar \omega$ and $U \ll \hbar \omega$. Furthermore,
\begin{align}\label{eq:actmol}
S&_{\mathrm{mol}}[b^{*},b] = \\ \nonumber
& \int_{0}^{\hbar\beta} d\tau \sum_{\rho,{\bf p}} b^{*}_{{\bf p},\rho}(\tau) \Bigg{\{}\hbar \frac{\partial}{\partial \tau} + \epsilon_{{\bf p}} + K_{\rho} - \mu_{\rho} \Bigg{\}}  b_{{\bf p},\rho}(\tau),
\end{align}
where $\epsilon_{{\bf p}} = \hbar^{2} {\bf p}^{2}/2 m_{\mathrm{d}}$. Finally, the part that describes the interaction between the photons and molecules can be rewritten as
\begin{align}\label{eq:actcoup}
&S_{\mathrm{c}}[a^{*}, a,b^{*}, b]= \\ \nonumber
&\frac{i}{\sqrt{2 A V}} \int_{0}^{\hbar\beta} d\tau  \sum_{i,{\bf k}, {\bf k}^{\prime},q}  \Big{\{} g_{{\bf k},{\bf k}^{\prime},i} a^{*}_{i}(\tau) b^{*}_{({\bf k}^{\prime},q),\downarrow}(\tau) \\ \nonumber
&\,\,\,\,\,\times \Big{(}b_{({\bf k}, q_{-}),\uparrow}(\tau) - b_{({\bf k}, q_{+}),\uparrow}(\tau) \Big{)} + g^{*}_{{\bf k},{\bf k}^{\prime},i} a_{i}(\tau) \\ \nonumber
&\,\,\,\,\,\times \left(b^{*}_{({\bf k}, q_{-}),\uparrow}(\tau) - b^{*}_{({\bf k}, q_{+}),\uparrow}(\tau) \right) b_{({\bf k}^{\prime},q),\downarrow}(\tau) \Big{\}},
\end{align} 
where ${\bf k}$ and ${\bf k}^{\prime}$ are two-dimensional and $q_{\pm} = q \pm k_{\gamma}$. Above and in the following, ${\bf p}$ is a three-dimensional momentum vector and ${\bf k}$ is two dimensional. Furthermore,
\begin{align}
g_{{\bf k},{\bf k}^{\prime},j} = g \int d{\bf x} \, w_{{\bf 0}}({\bf x} - {\bf x}_{j}) e^{i ({\bf k} - {\bf k}^{\prime}) \cdot {\bf x}},
\end{align}
and by using Eq.\,\eqref{eq:harmos}, 
\begin{align}\label{eq:coupli}
g_{{\bf k},{\bf k}^{\prime},j} &=  g_{\mathrm{m}} e^{i ({\bf k} - {\bf k}^{\prime}) \cdot {\bf x}_{j} - \hbar ({\bf k} - {\bf k}^{\prime})^{2} / 2 m \omega},
\end{align}
with $g_{\mathrm{m}} = \sqrt{4 g^{2} \pi \hbar / m \omega}$. The total action given by the sum of Eqs.\,(\ref{eq:photact}-\ref{eq:actcoup}), decribes the photon gas coupled to dye molecules in a periodic lattice in the single-band approximation. All the parameters that enter in this theory are now expressed into the experimentally tunable parameters $\lambda$ and $V_{0}$.

\section{Mean-field Theory}
\label{sec:simmod}
The model derived in the previous section is rather complicated due to the fixed longitudinal momentum $k_{\gamma}$ and the dependence of $g_{{\bf k},{\bf k}^{\prime},i}$ on two independent momenta. To get a better understanding of the physics involved in this system, we first consider a simplified model that neglects both the non-zero value of $k_{\gamma}$ and the non-diagonal coupling in $g_{{\bf k},{\bf k}^{\prime},i}$. We come back to the effects of these approximations in the second part of this section.

\subsection{Toy model}
We consider a toy model consisting of photons and molecules, where the molecules can only absorp and emit photons with very small momenta compared to the typical molecular momenta momentum. Thus, we consider a model with $S_{\mathrm{ph}}[a^{*}, a]$ and $S_{\mathrm{mol}}[b^{*},b]$ given by respectively Eqs.\,\eqref{eq:photact} and \eqref{eq:actmol}, and $S_{\mathrm{c}}[a^{*}, a,b^{*}, b]$ is changed into
\begin{align}\label{eq:BH}
S_{\mathrm{c}}&[a^{*}, a,b^{*}, b]= \frac{g_{\mathrm{m}}}{\sqrt{A} \sqrt{V}} \int_{0}^{\hbar\beta} d\tau \sum_{i,{\bf p}} \Big{\{} a^{*}_{i}(\tau) \\ \nonumber
&\times b^{*}_{{\bf p},\downarrow}(\tau) b_{{\bf p},\uparrow}(\tau) + a_{i}(\tau) b^{*}_{{\bf p},\uparrow}(\tau) b_{{\bf p},\downarrow}(\tau) \Big{\}}.
\end{align}
Now, we use a mean-field approach to calculate the phase diagram of the photons. Therefore, we introduce the order parameter $\psi  = \langle  a_{i}(\tau) \rangle$. However, due to the coupling with the molecules an expection value of $a_{i}(\tau)$ will also induce a non-zero value for $\phi_{{\bf p}} = \langle b^{*}_{{\bf p},\uparrow}(\tau) b_{{\bf p},\downarrow}(\tau) \rangle$. Physically this implies that the molecules are forced into a linear superposition of its internal states. In the language of magnetism this means that the pseudospin of the molecules gets also a component in the x-y plane. 
\newline
\indent In the end we are only interested in the photons, and therefore we want to calculate $\phi_{{\bf p}}$ as a function of $\psi$. Since $S_{\mathrm{ph}}[a^{*}, a]$ is irrelevant for this calculation, we first only consider $S_{\mathrm{c}}[a^{*}, a,b^{*}, b]$ and $S_{\mathrm{mol}}[b^{*},b]$. Up to linear order in the fluctuations, $S_{\mathrm{c}}[a^{*}, a,b^{*}, b]$ is given by
\begin{align}\label{eq:actmolmf}
S&_{\mathrm{c}}[a^{*}, a,b^{*}, b] = -2 \tilde{g}_{\mathrm{m}} N_{\mathrm{s}} \psi \int_{0}^{\hbar\beta} d\tau \sum_{{\bf p}} \phi_{{\bf p}}\\ \nonumber
&+ \tilde{g}_{\mathrm{m}} \int_{0}^{\hbar\beta} d\tau \sum_{i,{\bf p}}  \phi_{{\bf p}} (a_{i}(\tau) + a^{*}_{i}(\tau)) \\ \nonumber
&+ \tilde{g}_{\mathrm{m}} N_{\mathrm{s}} \psi \int_{0}^{\hbar\beta} d\tau \sum_{{\bf p}} \left\{ b^{*}_{{\bf p},\downarrow}(\tau) b_{{\bf p},\uparrow}(\tau) + b^{*}_{{\bf p},\uparrow}(\tau) b_{{\bf p},\downarrow}(\tau) \right\},
\end{align}
where $\tilde{g}_{\mathrm{m}} = g_{\mathrm{m}} / \sqrt{A} \sqrt{V}$, $N_{\mathrm{s}}$ is the number of lattice sites and without loss of generality we assumed that both $\psi$ and $\phi_{{\bf p}}$ are real. 
\newline
\indent We focus on the last part of the right-hand side of Eq.\,\eqref{eq:actmolmf} to obtain an expression for $\phi_{{\bf p}}$. By using a Matsubara expansion we can write for the part of the action that depends on the molecular fields $b_{\uparrow}$ and $b_{\downarrow}$,
\begin{align}
S_{\mathrm{mf}}&[b^{*}, b] = \sum_{{\bf p}, n}  u^{*}_{{\bf p},n}
\left [ \begin{array} {cc}
G^{-1}_{\uparrow} & \tilde{g}_{\mathrm{m}} N_{\mathrm{s}} \psi \\
\tilde{g}_{\mathrm{m}} N_{\mathrm{s}} \psi & G^{-1}_{\downarrow}
\end{array} \right ] u_{{\bf p},n},
\end{align}
where $G^{-1}_{\rho} = -i \hbar \omega_{n} + \epsilon_{{\bf p}} + K_{\rho} - \mu_{\rho}$ and
\begin{align}
u_{{\bf p},n} =
\left [ \begin{array} {c}
b_{{\bf p}, n, \uparrow} \\
b_{{\bf p}, n, \downarrow}
\end{array} \right ].
\end{align}
Now we perform a unitary transformation to diagonalize the action. Thus, we define
\begin{align}
v_{{\bf p},n} = \left [ \begin{array} {c}
\beta_{{\bf p}, n, \uparrow} \\
\beta_{{\bf p}, n, \downarrow}
\end{array} \right ]
=
\left [ \begin{array} {cc}
\cos(\theta) & -\sin(\theta) \\
\sin(\theta) & \cos(\theta)
\end{array} \right ]
\left [ \begin{array} {c}
b_{{\bf p}, n, \uparrow} \\
b_{{\bf p}, n, \downarrow}
\end{array} \right ].
\end{align}
Rewriting the action in terms of $v_{{\bf p},n}$ diagonalizes 
\begin{figure}[t]
\centerline{\includegraphics{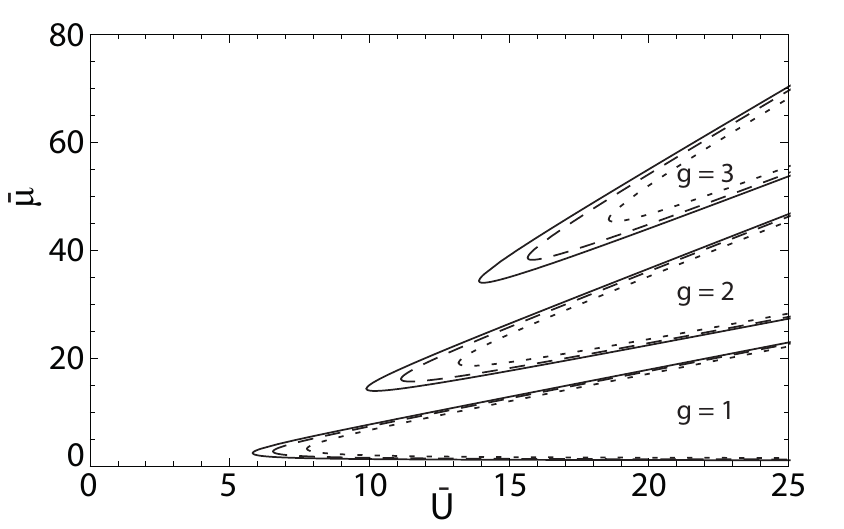}}
 \caption{Phase diagram of the Bose-Hubbard Hamiltonian with the photon-molecule interaction parameter $\gamma$. From bottom to top the three contributions are for respectively 1,2 or 3 particles per site. The solid, dashed and dotted line are for $\gamma = 0$, $\gamma = 1/2$ and $\gamma = 1$.}
 \label{fig:Motloops}
\end{figure}
the action with on the diagonal $-i \hbar \omega_{n} + \epsilon_{{\bf p}} + \Delta/2 - \lambda_{\pm}$ and
\begin{align}
\lambda_{\pm} = \Big{(}\mu_{\downarrow} + \mu_{\uparrow} \mp \sqrt{(\Delta - \Delta \mu)^{2} + 4 \tilde{g}_{\mathrm{m}}^{2} N_{\mathrm{s}}^{2} \psi^{2}} \Big{)}/2.
\end{align}
Moreover,
\begin{align}
\sin(2 \theta) &= \frac{2 \tilde{g}_{\mathrm{m}} N_{\mathrm{s}} \psi}{\sqrt{(\Delta - \Delta \mu)^{2} + 4 \tilde{g}_{\mathrm{m}}^{2} N_{\mathrm{s}}^{2} \psi^{2}}}, \\ \nonumber
\cos(2 \theta) &= \frac{\Delta - \Delta \mu}{\sqrt{(\Delta - \Delta \mu)^{2} + 4 \tilde{g}_{\mathrm{m}}^{2} N_{\mathrm{s}}^{2} \psi^{2}}}.
\end{align}
Here we defined $\Delta \mu = \mu_{\uparrow} - \mu_{\downarrow}$. Furthermore, since $\langle \beta^{*}_{\downarrow,{\bf p}}(\tau) \beta_{\uparrow,{\bf p}}(\tau) \rangle = \langle \beta_{\downarrow,{\bf p}}(\tau) \beta^{*}_{\uparrow,{\bf p}}(\tau) \rangle = 0$, we obtain
\begin{align}\label{eq:expmol}
\phi_{{\bf p}} &= \langle b^{*}_{{\bf p},\downarrow}(\tau) b_{{\bf p},\uparrow}(\tau) \rangle \\ \nonumber
&= \frac{\tilde{g}_{\mathrm{m}} N_{\mathrm{s}} \psi}{\sqrt{(\Delta - \Delta \mu)^{2} + 4 \tilde{g}_{\mathrm{m}}^{2} N_{\mathrm{s}}^{2} \psi^{2}}} \\ \nonumber
&\times \left\{N_{\mathrm{MB}}(\epsilon_{{\bf p}} + \Delta/2 - \lambda_{+}) -  N_{\mathrm{MB}}(\epsilon_{{\bf p}} +\Delta/2 - \lambda_{-}) \right\},
\end{align}
where we again considered the Maxwell-Boltzmann limit as the system is at room temperature. As already mentioned before, this equation explicitly shows that $\psi \neq 0$ also implies a non-zero value of $\phi_{{\bf p}}$. 
\newline
\indent In order to further investigate the properties of the photons, we substitute this result into the action in Eq.\,\eqref{eq:actmolmf}. To compare with results for standard Hubbard models as in for example Ref.\,\cite{Dries}, we switch to the Hamiltonian formalism. In the thermodynamic limit the effective Hamiltonian that describes the photons in the mean-field approximation is now given by
\begin{align}
\hat{H}^{\mathrm{eff}} &= z t \psi^{2} N_{\mathrm{s}} \left[1 + 2 \gamma \right] + \frac{U}{2} \sum_{i} \hat{n}_{i} \left(\hat{n}_{i} - 1 \right) \\ \nonumber
&- z t \psi \left[1 + \gamma \right] \sum_{i} \left(\hat{a}_{i} + \hat{a}^{\dagger}_{i} \right) - \mu \sum_{i} \hat{n}_{i},
\end{align}
where we only considered tunneling between nearest neighbours, $z$ is the number of nearest neighbours, $\hat{n}_{i} = \hat{a}_{i}^{\dagger} \hat{a}_{i}$ is the photon-number operator and
\begin{align}\label{eq:gamma}
\gamma &= \frac{\tilde{g}_{\mathrm{m}}}{z t} \sum_{{\bf p}} \phi_{{\bf p}}\\ \nonumber
&= \frac{g_{\mathrm{m}}^{2}}{z t} \frac{P(\Delta \mu)}{\Delta \mu - \Delta} \left(\frac{m_{\mathrm{d}}}{m_{\mathrm{d,real}}} \right)^{3/2} n_{\mathrm{s}} n_{\mathrm{mol}}.
\end{align}
Here $P(\Delta \mu)$ denotes the polarization of the molecules as defined in Ref.\,\cite{AW}, $m_{\mathrm{d,real}}$ is the real mass of the dye molecules and $n_{\mathrm{mol}}$ equals the density of molecules. Furthermore, $n_{\mathrm{s}} = N_{\mathrm{s}} / A$ is the density of sites. Moreover, note that $\gamma > 0$ as the polarization $P(\Delta\mu)$ and $\Delta\mu - \Delta$ have the same sign for all values of $\Delta\mu$. Moreover, recall that we modeled the molecules as a two-level system with an effective mass, and therefore the ratio $m_{\mathrm{d}}/m_{\mathrm{d,real}}$ appears in the final result. 
\newline
\indent By introducting $\bar{U} = U / zt$ and $\bar{\mu} = \mu / zt$, we define $\hat{H}^{\mathrm{eff}} = z t \sum_{i} \hat{H}_{i}$ and the Hamiltonian $\hat{H}_{i}$ at each site $i$ as
\begin{align}
\hat{H}_{i} &= \frac{\bar{U}}{2} \hat{n}_{i} \left(\hat{n}_{i} - 1 \right)  + \psi^{2} \left[1 + 2 \gamma \right] \\ \nonumber
&- \bar{\mu} \hat{n}_{i} - \psi \left[1 + \gamma \right] \left(\hat{a}_{i} + \hat{a}^{\dagger}_{i} \right).
\end{align}
To calculate the phase diagram of the photons in the dye-filled cavity, we use the usual Landau theory for second-order phase transitions. Hence, we write the energy of the ground state as
\begin{align}
E_{\mathrm{g}}(\psi) = a_{0}(g,\bar{U}, \bar{\mu}) + a_{2}(g,\bar{U}, \bar{\mu}) \psi^{2} + \mathcal{O}(\psi^{4}),
\end{align}
and we minimize this energy. The corresponding value of $\psi$ determines the phase of the system. For $a_{2}(g,\bar{U}, \bar{\mu}) \geq 0$, $\psi = 0$  and the system is inside a Mott lobe. Differently, for $a_{2}(g,\bar{U}, \bar{\mu}) < 0$ we have that $\psi \neq 0$ and the photons are in the superfluid phase. This distinction becomes clear if we consider the ground state of the system. At the level of mean-field theory, the ground state inside the Mott lobe is a state with a well-defined number of particles at each site. Therefore we obtain $\psi = \langle \hat{a}_{i} \rangle = 0$ as the states with different number of photons per site are orthogonal. Furthermore, in the superfluid phase the number of particles per site is not sharply defined. Thus, the ground state is a linear superposition of number states and therefore $\psi \neq 0$. 
\newline
\indent To investigate the phase transition we determine $a_{2}(g,\bar{U}, \bar{\mu})$ by following Ref.\,\cite{Dries} and perform second-order perturbation theory. Therefore, we split the Hamiltonian in an exactly solvable part $\hat{H}_{0}$ and a pertubation $\psi \left[1 + \gamma \right]  \hat{V}$. In this case 
\begin{align}\label{eq:ham}
\hat{H}_{0} = \frac{\bar{U}}{2} \hat{n}_{i} \left(\hat{n}_{i} - 1 \right)  + \psi^{2} \left[1 + 2 \gamma \right] - \bar{\mu} \hat{n}_{i},
\end{align}
and
\begin{align}
\hat{V} = \left(\hat{a}_{i} + \hat{a}^{\dagger}_{i} \right).
\end{align}
In pertubation theory the second-order correction to the energy reads
\begin{align}
E^{(2)}_{\mathrm{g}} = \psi^{2}\left[1 + \gamma \right]^{2} \sum_{n \neq g} \frac{|\langle n |\hat{V}| g \rangle|^{2}}{E^{(0)}_{\mathrm{g}} - E^{(0)}_{\mathrm{n}}},
\end{align}
where $|n\rangle$ is the state with $n$ particles and for $n = g$ this state is the ground state. Furthermore, $E^{(0)}_{\mathrm{n}}$ is the energy of the state $|n\rangle$ with respect to the exactly solvable Hamiltonian $\hat{H}_{0}$. Again analogous to Ref.\,\cite{Dries}, we obtain
\begin{align}\label{eq:coef1}
a_{2}&(g,\bar{U}, \bar{\mu}) = \\ \nonumber
& \left[1 + \gamma \right]^{2} \left(\frac{g}{\bar{U} (g -1) - \bar{\mu}} + \frac{g + 1}{\bar{\mu} - \bar{U} g} \right) + 1 + 2 \gamma
\end{align}
for $\bar{U} (g-1) < \bar{\mu} < \bar{U} g$. The boundary of the Mott lobes can be found by solving $a_{2}(g,\bar{U}, \bar{\mu}_{\pm}) = 0$ for $\bar{\mu}_{\pm}$. We find
\begin{align}\label{eq:mumot}
\bar{\mu}_{\pm} &= \frac{1}{2} \left((2g - 1) \bar{U} - 1/\Gamma \right) \\ \nonumber
&\pm \frac{1}{2 \Gamma} \sqrt{1 + \bar{U} \Gamma (\bar{U} \Gamma - 4 g - 2)},
\end{align}
where $\Gamma = (1 + 2\gamma)/(1 + \gamma)^{2} < 1$. In Fig.\,\ref{fig:Motloops} we present a plot of the two branches of Eq.\,\eqref{eq:mumot} for several values of $\gamma$. We obtain that for increasing $\gamma$, the size of the Mott lobes decreases. Physically this makes sense as the absorption and emision of photons by the molecules effectively increases the  hopping of photons between different lattice sites. Furthermore, we note that for increasing number of particles in the ground state the effect of $\gamma$ also increases. Moreover, the smallest $\bar{U}$ for each Mott lobe is equal to
\begin{align}
\bar{U}_{\mathrm{c}} = \left(2g+1 + \sqrt{(2g + 1)^{2}-1}\right)/\Gamma = \bar{U}_{\mathrm{c}, 0}/\Gamma,
\end{align}
with $\bar{U}_{\mathrm{c}, 0}$ the result obtained in Ref.\,\cite{Dries} for $\gamma = 0$ and correspondingly $\Gamma = 1$. 
\newline
\indent To estimate the value of $\gamma$, we want to express this coefficient in terms of the experimental parameters $\lambda$ and $V_{0}$. As mentioned before, $\lambda$ is twice the lattice spacing and $V_{0}$ is the depth of the lattice potential. 
\begin{figure}[t]
\centerline{\includegraphics[scale=1]{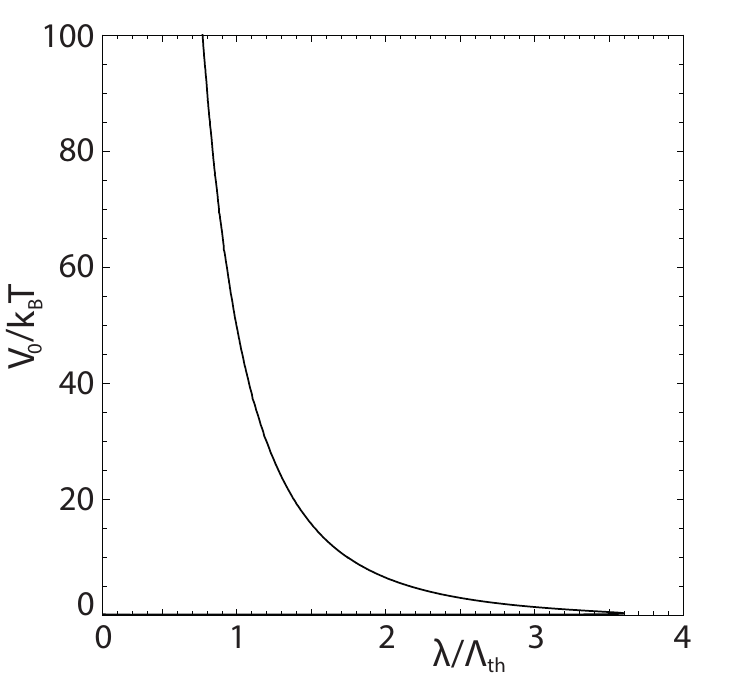}}
 \caption{A plot of the values of $\gamma$ in terms of the experimental parameters $V_{0}$ and $\lambda$. We used $n_{\mathrm{mol}} = 9 \cdot 10^{23}\,\mathrm{m}^{-3}$, $T = 300\,\mathrm{K}$ and $\beta |\Delta \mu - \Delta| < 1$. Furthermore, $\Lambda_{\mathrm{th}}$ is the thermal de Broglie wavelength. On the right-hand side of the solid line $\gamma > 1/2$ and the corrections to the Mott lobes are important. To the left of the solid line $\gamma < 1/2$ and the effect of the photon-molecule coupling $\gamma$ on the Mott lobes is small. Finally, these results are only valid if thermal fluctuations are small $k_{\mathrm{B}} T \ll \hbar \omega$ or $4 \pi V_{0}/k_{\mathrm{B}}T \gg (\lambda/\Lambda_{\mathrm{th}})^{2}$.}
 \label{fig:Gammaplot}
\end{figure}
From Eq.\,\eqref{eq:gamma} it follows that we have to find an expression for $t$. From Ref.\,\cite{tunneling}, we obtain
\begin{align}
\tilde{t} = \frac{4 \tilde{V}_{0}^{3/4}}{\pi^{1/4}} \left(\frac{\Lambda_{\mathrm{th}}}{\lambda}\right)^{1/2} \exp\left\{-2 \left(\frac{\lambda}{\Lambda_{\mathrm{th}}} \right) \sqrt{\frac{\tilde{V}_{0}}{\pi}} \right\}.
\end{align}
Here $\Lambda_{\mathrm{th}} = (2 \pi \hbar^{2}/m k_{\mathrm{B}} T)^{1/2}$ is the thermal de Broglie wavelength, $\tilde{V}_{0} = V_{0}/k_{\mathrm{B}} T$ and $\tilde{t} = t/k_{\mathrm{B}} T$. For the two-dimensional squared periodic potential considered here, we take $n_{\mathrm{s}} = 4 /\lambda^{2}$ and $z =4$. Hence, 
\begin{align}
\gamma &= \frac{1}{4 \pi^{1/4}} \sqrt{\frac{\Lambda_{\mathrm{th}}}{\lambda}} \frac{1}{\tilde{V}_{0}^{5/4}}  \exp\left\{2 \left(\frac{\lambda}{\Lambda_{\mathrm{th}}} \right) \sqrt{\frac{\tilde{V}_{0}}{\pi}} \right\} \\ \nonumber
&\times \frac{g^{2} \beta^{2}P(\Delta \mu)}{\beta (\Delta \mu - \Delta)} \left(\frac{m_{\mathrm{eff}}}{m_{\mathrm{real}}} \right)^{3/2} n_{\mathrm{mol}}.
\end{align}
In Fig.\,\ref{fig:Gammaplot} we use this expression for $\gamma$ to illustrate the phase diagram for $\beta |\Delta \mu - \Delta| < 1$ and $z = 4$. For these chemical potentials we can approximate $P(\Delta \mu) / \beta (\Delta \mu - \Delta) \simeq 1/2$ and for other quantites such as $g$, $m_{\mathrm{d}}$ and $m_{\mathrm{d,real}}$ we take the numerical values as found in Ref.\,\cite{AW}.

\subsection{Photon Model}
Up to now we only considered the terms with diagonal coupling in momentum space and we neglected the fixed momentum $k_{\gamma}$ of the photon in the longitudional direction. In the following we consider the effect of these approximations on the phase diagram presented in the previous section. Therefore, here we consider the full action consisting of the sum of Eqs.\,(\ref{eq:photact}-\ref{eq:actcoup}). In accordance with the previous calculation, we define $\langle b^{*}_{({\bf k}^{\prime}, q),\downarrow}(\tau) b_{({\bf k}, q_{-}),\uparrow}(\tau) \rangle = \phi_{{\bf k},{\bf k}^{\prime},q_{-}}$ and $\langle a_{\mathrm{i}} \rangle = \psi$. The expectation value of the molecular fields that depend on $q_{+}$ are denoted by $\phi_{{\bf k},{\bf k}^{\prime},q_{+}}$. 
\newline
\indent Now we perform a mean-field approximation and we calculate $\phi_{{\bf k},{\bf k}^{\prime},p_{\mathrm{z}}}$ as a function of $\psi$. Therefore, we consider $S_{\mathrm{mol}}[b^{*},b]$ and
\begin{align}
&S_{\mathrm{c}}[a^{*}, a,b^{*}, b] = \frac{i N_{\mathrm{s}} \psi}{\sqrt{2 A V}} \int_{0}^{\hbar\beta} d\tau \, \sum_{{\bf k},{\bf G},q} g_{{\bf G}} \\ \nonumber
&\times \Big{\{}b^{*}_{({\bf k}+{\bf G},q),\downarrow}(\tau) ( b_{({\bf k}, q_{-}),\uparrow}(\tau)  - b_{({\bf k}, q_{+}),\uparrow}(\tau)) \\ \nonumber
&+ ( b^{*}_{({\bf k}, q_{-}),\uparrow}(\tau)  - b^{*}_{({\bf k}, q_{+}),\uparrow}(\tau)) b_{({\bf k}+{\bf G},q),\downarrow}(\tau)\Big{\}}, 
\end{align}
where without loss of generality we again assumed that $\psi$ is real. Furthermore, we performed the summation over the lattice sites. After this summation the coupling constant $g_{{\bf k},{\bf k}^{\prime},i}$ given by Eq.\,\eqref{eq:coupli}, only depends on a two-dimensional reciprocal lattice vector ${\bf G}$. Here we consider Eq.\,\eqref{eq:pot}, and therefore we have a cubic lattice with lattices sites $(n \lambda/2 - \lambda/4,m \lambda/2 - \lambda/4)$ with $n$ and $m$ integers. Therefore, ${\bf G} = 4 \pi {\bf n} / \lambda$ with ${\bf n} = (n_{x}, n_{y})$ and $n_{x}$ and $n_{y}$ integers. Hence,
\begin{align}
g_{{\bf G}} = g_{\mathrm{m}} \exp\left\{- \frac{2}{\tilde{\lambda}} \sqrt{\frac{\pi}{\tilde{V_{0}}}} \left(n_{x}^{2} + n_{y}^{2} \right) \right\},
\end{align}
where $\tilde{\lambda} = \lambda/\Lambda_{\mathrm{th}}$ and $\tilde{V}_{0} = V_{0} / k_{\mathrm{B}}T$.  
\newline
\indent To calculate $\phi_{{\bf k},{\bf k}^{\prime},q}$ we in principle have to invert an infinite-dimensional matrix. However, we are generally interested in the phase transition and therefore we only need to calculate the inverse of this infinite-dimensional matrix up to linear order in $\psi$. Since all the off-diagonal terms of this infinite dimensional matrix are already linear in $\psi$, we obtain up to linear order in $\psi$
\begin{align}
&\phi_{{\bf k},{\bf k} + {\bf G},q_{+}} = \frac{-i N_{\mathrm{s}} \psi g_{{\bf G}}}{\sqrt{2 A V}}\\ \nonumber
&\times \frac{N_{\mathrm{MB}}(\epsilon({\bf k},q_{+}) - \mu_{\uparrow} + \Delta) - N_{\mathrm{MB}}(\epsilon({\bf k} + {\bf G},q) - \mu_{\downarrow})}{\epsilon({\bf k},q_{+}) - \epsilon({\bf k} + {\bf G},q) + \Delta - \Delta\mu},
\end{align}
where we considered the classical limit. 
\begin{figure}[t]
\centerline{\includegraphics[scale=1]{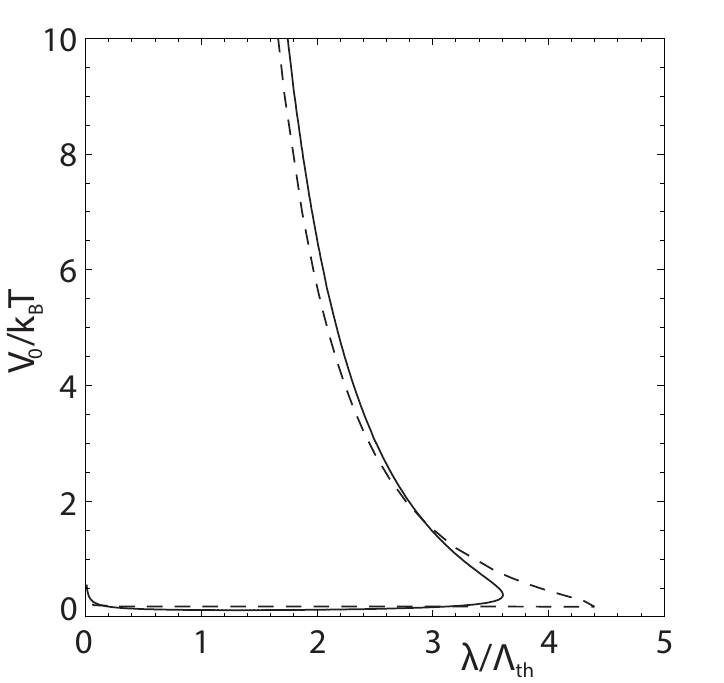}}
 \caption{A plot of the values of $\gamma$ for the complete photon model and the simplified model in terms of the experimental parameters $V_{0}$ and $\lambda$. The simplified model is represented by the solid line and the full model is depicted by the dashed line. On the right-hand side of both lines $\gamma > 1/2$ and here the Mott lobes are noticeably affected by the photon-molecule coupling. In this plot, we take $\Lambda_{\mathrm{th}} \simeq 1.6\cdot 10^{-6}\,\mathrm{m}$ and the numerical values for the other parameters and the conditions are the same as denoted below Fig.\,\ref{fig:Gammaplot}.}
 \label{fig:PDFM}
\end{figure}
We again combine the effects of the molecules in the parameter $\gamma$ which in this case reads
\begin{align}
\gamma &= \frac{n_{\mathrm{s}}}{z t V} \sum_{{\bf k},n_{x},n_{y},q} g_{\mathrm{m}}^{2} \exp\left\{- \frac{4}{\tilde{\lambda}} \sqrt{\frac{\pi}{\tilde{V_{0}}}} \left(n_{x}^{2} + n_{y}^{2} \right) \right\} \\ \nonumber
&\times  \frac{N_{\mathrm{MB}}(\epsilon({\bf k},q_{+}) - \mu_{\uparrow} + \Delta) - N_{\mathrm{MB}}(\epsilon({\bf k} + {\bf G},q) - \mu_{\downarrow})}{\Delta\mu - \Delta +  \epsilon({\bf k} + {\bf G},q) - \epsilon({\bf k},q_{+})},
\end{align}
where we used that the contributions of both $\phi_{{\bf k},{\bf k} + {\bf G},q_{+}}$ and $\phi_{{\bf k},{\bf k} + {\bf G},q_{-}}$ are the same. Note that for $k_{\gamma} = 0$ and if we only consider $n_{x} = n_{y} = 0$, we recover our previous result from the simplified model. 
\newline
\indent To determine $\gamma$ we have to evaluate a sum over an infinite number of terms. However, by performing a numerical analysis we obtain that the contribution rapidly decreases for increasing $n_{x}$ and $n_{y}$. Therefore, it suffices to only take into account the terms where both $n_{x} \leq 1$ and $n_{y} \leq 1$. Since $\gamma$ depends on $\Delta \mu$, we obtain a different phase diagram for every value of $\Delta \mu$. However, although there are quantitative differences, the qualitative results of the calculations are not dependent on the precise value of $\Delta \mu$. Therefore, in comparison with the previous calculation we again take $\beta |\Delta - \Delta \mu| < 1$.
\newline
\indent In Fig.\,\ref{fig:PDFM} we show a comparison between the values of $\gamma$ in the thermodynamic limit as a function of the experimental parameters $V_{0}$ and $\lambda$, for the full photon model discussed here and the simplified model used previously. As can be seen from the figure, both results are qualitatively the same. However, quantitatively there are some differences between both results. The largest difference occurs for relatively small values of $V_{0}$. Namely, for $V_{0}/ k_{\mathrm{B}}T < 2$ the full photon model results into a larger region where the value of $\gamma$ is relatively small. For larger $V_{0}$ both results are remarkably close, and the full model has only a marginally larger region where the effects of the molecule coupling are important. 

\section{Number fluctuations inside the Mott lobes}
\label{sec:numfluc}
In the previous section we constructed a self-consistent mean-field theory to investigate the effect of the photon-molecule coupling on the phase diagram of the photonic lattice in a dye-filled optical microcavity. However, to obtain more information about the physics in the Mott lobes, we have to go beyond mean-field theory and also include fluctuations. Since the theory in the previous section can not easily be generalised to describe these fluctuations, we now use functional methods instead of the operator methods used previously. This has the advantage that it is relatively straightforward to include fluctuations at zero temperature. 
\newline
\indent Similar to the approach in Ref.\,\cite{AW}, we integrate out the molecules and expand up to second order in the coupling constant $g$. Thus, we obtain
\begin{align}\label{eq:action}
S&[a^{*}, a] = \int_{0}^{\hbar\beta} d\tau \sum_{i} a^{*}_{i}(\tau) \left(\hbar \frac{\partial}{\partial \tau} - \mu \right) a_{i}(\tau) \\ \nonumber
&+ \frac{U}{2} \int_{0}^{\hbar\beta} d\tau \sum_{i} a^{*}_{i}(\tau) a^{*}_{i}(\tau) a_{i}(\tau) a_{i}(\tau) \\ \nonumber
&- \int_{0}^{\hbar\beta} d\tau \int d\tau^{\prime} \sum_{i,j} a^{*}_{i}(\tau)  G_{i,j}^{-1}(\tau - \tau^{\prime}) a_{j}(\tau ^{\prime}),
\end{align}
with
\begin{align}\label{eq:greenmot}
G_{i,j}^{-1}(\tau - \tau^{\prime}) = t_{i,j} \delta(\tau - \tau^{\prime}) - \hbar\Sigma_{i,j}(\tau - \tau^{\prime}),
\end{align}
and
\begin{align}\label{eq:selfmot}
\hbar\Sigma_{i,j}&(\tau - \tau^{\prime}) = \frac{1}{A V} \sum_{{\bf k},{\bf k}^{\prime},p_{z}} g_{{\bf k},{\bf k}^{\prime},i} g^{*}_{{\bf k},{\bf k}^{\prime},j} \\ \nonumber
&\times G_{\downarrow}(\tau^{\prime}-\tau, {\bf k}^{\prime}, p_{z}) G_{\uparrow}(\tau-\tau^{\prime}, {\bf k}, p_{z} + k_{\gamma}).
\end{align}
Here, the coupling constant $g_{{\bf k},{\bf k}^{\prime},i}$ is defined in Eq.\,\eqref{eq:coupli} and $ G_{\sigma}(\tau^{\prime}-\tau, {\bf k}^{\prime}, p_{z})$ denotes the Green's function of the excited or ground-state molecules. From these expressions we obtain that the interaction with the molecules is an additional mechanism for the photons to hop between different lattice sites. Since $\hbar\Sigma_{i,i}(\tau - \tau^{\prime}) \neq 0$, the interaction with the molecules also contributes to the on-site Green's function of the photons. To decouple the hopping term we perform a Hubbard-Stratonovich transformation to the action, and we write
\begin{align}\label{eq:actexp1}
S&[a,a^{*},\psi^{*},\psi] = S[a^{*}, a] + \int_{0}^{\hbar\beta} d\tau \int_{0}^{\hbar\beta} d\tau^{\prime} \\ \nonumber
&\times \sum_{i,j} \left(a_{i}^{*}(\tau) - \psi_{i}^{*}(\tau) \right) G_{i,j}^{-1}(\tau - \tau^{\prime}) \left(a_{j}(\tau^{\prime}) - \psi_{j}(\tau^{\prime}) \right),
\end{align}
where $\psi_{i}(\tau)$ is the complex order-parameter field. Following the procedure described in Ref.\,\cite{Dries}, we calculate the action up to second order in $\psi$ and obtain
\begin{align}\label{eq:action2}
S&^{(2)}[\psi^{*}, \psi] =  \\ \nonumber
& \int_{0}^{\hbar\beta} d\tau \int_{0}^{\hbar\beta} d\tau^{\prime}\, \sum_{i,j} \psi_{i}^{*}(\tau)  G_{i,j}^{-1}(\tau - \tau^{\prime})  \psi_{j}(\tau^{\prime})\\ \nonumber
& - \frac{1}{\hbar} \int_{0}^{\hbar\beta} d\tau \int_{0}^{\hbar\beta} d\tau^{\prime} \int_{0}^{\hbar\beta} d\tau^{\prime\prime} \int_{0}^{\hbar\beta} d\tau^{\prime\prime\prime}\,\sum_{i,j,i^{\prime},j^{\prime}} \psi_{i}^{*}(\tau) \\ \nonumber
&\times G_{i,j^{\prime}}^{-1}(\tau - \tau^{\prime\prime\prime}) \langle a_{j^{\prime}}(\tau^{\prime\prime\prime}) a_{i^{\prime}}^{*}(\tau^{\prime\prime})\rangle_{0} G_{i^{\prime},j}^{-1}(\tau^{\prime\prime} - \tau^{\prime}) \psi_{j}(\tau^{\prime}).
\end{align}
Here $\langle ... \rangle_{0}$ denotes the expectation value with respect to the action in Eq.\,\eqref{eq:actexp1} for $G_{i,j}^{-1}(\tau - \tau^{\prime}) = 0$. Now, we separately calculate the first and second part of this action. First, we define
\begin{align}\label{eq:fieldmot}
\psi_{i}(\tau) = \frac{1}{\sqrt{\hbar \beta N_{\mathrm{s}}}} \sum_{{\bf k},n} \psi_{{\bf k},n} e^{i ({\bf k} \cdot {\bf x}_{i} - \omega_{n} \tau)},
\end{align}
where ${\bf k}$ only runs over the first Brillouin zone. Now, we substitute the Fourier expansions of $\psi_{i}(\tau)$ and $\hbar\Sigma_{i,j}(\tau - \tau^{\prime})$ and again we take $t_{ij}$ equal to $t$ for nearest neighbours and zero otherwhise. By performing the integrations over $\tau$ and $\tau^{\prime}$, we obtain
\begin{align}\label{eq:termmot}
&\int_{0}^{\hbar\beta} d\tau \int_{0}^{\hbar\beta} d\tau^{\prime}\, \sum_{i,j} \psi_{i}^{*}(\tau)  G_{i,j}^{-1}(\tau - \tau^{\prime})  \psi_{j}(\tau^{\prime}) \\ \nonumber
&=-\sum_{{\bf k},n} G^{-1}_{\mathrm{m}}({\bf k}, i \omega_{n}) \psi^{*}_{{\bf k},n}  \psi_{{\bf k},n},
\end{align}
where
\begin{align}
G^{-1}_{\mathrm{m}}({\bf k}, i \omega_{n}) = \epsilon_{{\bf k}} + \frac{16 \pi \hbar}{m \lambda^{2} \omega} \hbar\Sigma({\bf k},k_{\gamma},i \omega_{n}).
\end{align}
Here, we used that we are inside a Mott lobe where $\omega$ is sufficiently large and therefore we used the approximation that $|g_{{\bf k},{\bf k}^{\prime},i}|^{2} = |g_{\mathrm{m}}|^{2}$. Note that if we perform the Wick rotation to real frequencies in $\hbar\Sigma({\bf k},k_{\gamma},i \omega_{n})$, we obtain the retarded self-energy as calculated in Ref.\,\cite{AW}. 
Furthermore, the lattice dispersion $\epsilon_{{\bf k}}$ is given by
\begin{align}
\epsilon_{{\bf k}} = -2 t \sum_{j=1}^{2} \cos(k_{j} \lambda / 2). 
\end{align}
Similar to the calculation in the previous section, the periodicity of the photons induces the introduction of reciprocal lattice vectors for the molecules. The incorporation of these vectors gives additional contributions that are off-diagonal in momentum space. However, especially inside the Mott lobes these contributions are small. Since in this section we only consider that part of the phase diagram, these contributions are neglected throughout the rest of this section.
\newline
\indent The calculation of the second term of Eq.\,\eqref{eq:action2} is more involved. However, we only consider the zero-temperature case and therefore we can use some results from Ref.\,\cite{Dries}. Similarly, we obtain
\begin{align}
\langle a_{j^{\prime}}(\tau) a_{i^{\prime}}^{*}(\tau^{\prime})\rangle_{0} = \delta_{i^{\prime},j^{\prime}} \langle a_{i^{\prime}}(\tau) a_{i^{\prime}}^{*}(\tau^{\prime})\rangle_{0},
\end{align}
with
\begin{align}\label{eq:expecmot}
\langle a_{i}(\tau) a_{i}^{*}(\tau^{\prime})\rangle_{0} &= \Theta(\tau - \tau^{\prime}) (g+1) e^{(\mu - gU) (\tau - \tau^{\prime})/\hbar} \\ \nonumber
&+\Theta(\tau^{\prime} - \tau) g e^{(\mu - (g-1)U) (\tau - \tau^{\prime})/\hbar}.
\end{align}
Again, by only taking into account nearest-neighbour hopping we can evaluate the second term of Eq.\,\eqref{eq:action2} explicitly. By using Eqs.\,\eqref{eq:greenmot}, \eqref{eq:selfmot}, \eqref{eq:fieldmot} and \eqref{eq:expecmot}, we can perform the integration over imaginary time and summations over lattice sites. Combining this result with Eq.\,\eqref{eq:termmot}, we find
\begin{align}\label{eq:acpsi}
S^{(2)}[\psi^{*},\psi] = -\hbar \sum_{{\bf k},n} \psi^{*}_{{\bf k},n} G^{-1}({\bf k},i \omega_{n}) \psi_{{\bf k},n},
\end{align}
where the inverse Green's function obeys
\begin{align}\label{eq:greensplit}
\hbar &G^{-1}({\bf k},i \omega_{n}) = G^{-1}_{\mathrm{m}}({\bf k}, i \omega_{n}) \Bigg{\{} 1 + G^{-1}_{\mathrm{m}}({\bf k}, i \omega_{n})\\ \nonumber
&\times \left(\frac{g + 1}{-i\hbar\omega_{n} - \mu + g U} + \frac{g}{i\hbar\omega_{n} + \mu - (g-1) U}\right)\Bigg{\}}.
\end{align}
As the quadratic coefficient $a_{2}(g,\mu,U)$ of the Landau free energy coincides with $-\hbar G^{-1}({\bf 0},0) / z t$, this expression allows us to compare this theory with the result from the previous section. By using the expression for the self-energy in Ref.\,\cite{AW}, we obtain 
\begin{align}\label{eq:Land2}
-&\hbar G^{-1}({\bf 0},0) / z t = \\ \nonumber
& \left(1 + \gamma \right) \left[1 + \left(1 + \gamma \right) \left\{\frac{g}{\bar{U} (g -1) - \bar{\mu}} + \frac{g + 1}{\bar{\mu} - \bar{U} g}\right\} \right].
\end{align}
Since in this approach we do not take into account the non-zero expectation value of $\langle b^{*}_{{\bf p}^{\prime},\downarrow}(\tau) b_{{\bf p},\uparrow}(\tau) \rangle$, we arrive at a slightly different coefficient $a_{2}(g,\mu,U)$ as found in Eq.\,\eqref{eq:coef1}. Both mean-field theories are however qualitatively the same as they lead to shrinking Mott lobes for increasing values for $\gamma$. Although the exact position of the phase boundary of the Mott lobes is slightly different, we expect that inside these loops both approaches are equivalent as in this region of the phase diagram the expectation value of the photon annihilation operator and therefore $\langle b^{*}_{{\bf p}^{\prime},\downarrow}(\tau) b_{{\bf p},\uparrow}(\tau) \rangle$ is zero.

\subsection{Quasiparticle excitations}
First, we use the theory that includes the Gaussian fluctuations to obtain $\langle a_{i}(\tau) a_{i^{\prime}}^{*}(\tau^{\prime})\rangle$, and thereby calculating the quasihole and quasiparticle excitations in the Mott lobes. 
\begin{figure}[t]
\centerline{\includegraphics[scale=1]{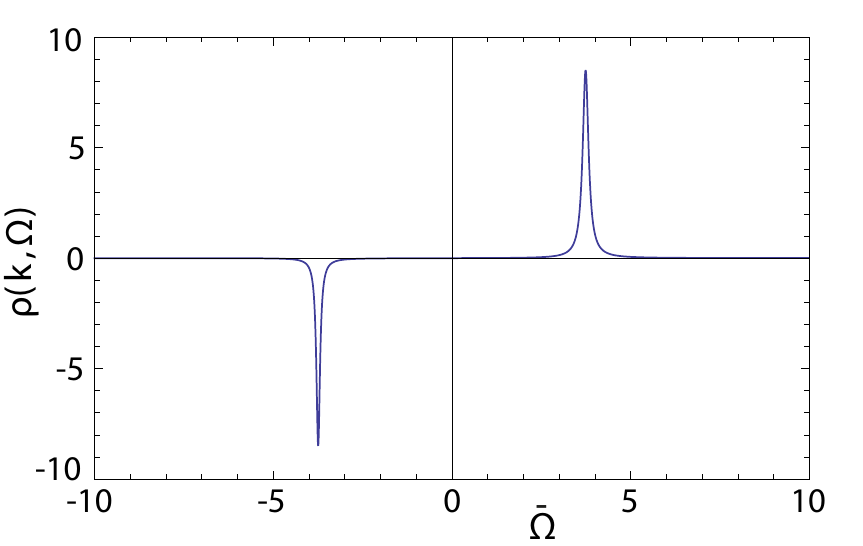}}
 \caption{A plot of the spectral function of the photons $\rho({\bf k},\Omega)$ inside the $g=1$ Mott lobe as a function of $\bar{\Omega} = \Omega/zt$ for ${\bf k} = 0$, $U /zt = 11$, $\mu / zt = 5$ and $\alpha_{\mathrm{lat}} = 10^{-2}$. The incorporation of the photon-molecule coupling broadens the peaks that are located at the quasihole and quasiparticle excitations.}
 \label{fig:SpecFM}
\end{figure}
To obtain a relation between this correlator of photon operators and the correlator of the Hubbard-Stratonovich fields $\langle \psi^{*}_{i}(\tau) \psi_{j}(\tau^{\prime}) \rangle$, we add sources $J_{i}(\tau)$ and $J^{*}_{i}(\tau)$ that couple to $a_{i}(\tau)$ and $a^{*}_{i}(\tau)$. Instead of the Hubbard-Stratonovich transformation used in Eq.\,\eqref{eq:actexp1}, we now add 
\begin{align}\label{eq:HS}
\int_{0}^{\hbar\beta} d\tau &\int_{0}^{\hbar\beta} d\tau^{\prime} \sum_{i,j} \Big{(}a_{i}^{*}(\tau) - \psi_{i}^{*}(\tau) + \left[J \cdot G\right]_{i}(\tau)\\ \nonumber
&\times G_{i,j}^{-1}(\tau - \tau^{\prime}) \Big{(} a_{j}(\tau^{\prime}) - \psi_{j}(\tau^{\prime}) \left[G \cdot J\right]_{j}(\tau^{\prime}) \Big{)},
\end{align}
to the action in Eq.\,\eqref{eq:action}. Here we introduced a short-hand notation for the convolution
\begin{align}
\left[J \cdot G\right]_{i}(\tau) = \sum_{i^{\prime}} \int d\tau^{\prime\prime}\,\hbar J^{*}_{i^{\prime}}(\tau^{\prime\prime}) G_{i^{\prime},i}(\tau^{\prime\prime} - \tau).
\end{align}
By differentiation of the partition function with respect to the currents, we obtain 
\begin{align}\label{eq:greencur}
\langle a^{*}_{i}(\tau) a_{j}(\tau^{\prime}) \rangle = \langle \psi^{*}_{i}(\tau) \psi_{j}(\tau^{\prime}) \rangle - \hbar G_{i,j}(\tau - \tau^{\prime}).
\end{align}
Denoting the Fourier transform of $\langle a^{*}_{i}(\tau) a_{j}(\tau^{\prime}) \rangle$ by $G_{\mathrm{ph}}({\bf k},i \omega_{n})$, we therefore find that
\begin{align}\label{eq:greensphot}
&-\frac{1}{\hbar} G_{\mathrm{ph}}({\bf k},i \omega_{n}) = \Bigg{(}\frac{g + 1}{-i\hbar\omega_{n} - \mu + g U} \\ \nonumber
&+ \frac{g}{i\hbar\omega_{n} + \mu - (g-1) U}\Bigg{)} \Bigg{\{} 1 + G^{-1}_{\mathrm{m}}({\bf k}, i \omega_{n})\\ \nonumber
&\times \left(\frac{g + 1}{-i\hbar\omega_{n} - \mu + g U} + \frac{g}{i\hbar\omega_{n} + \mu - (g-1) U}\right)\Bigg{\}}^{-1}.
\end{align}
The excitations of the photons in the Mott lobes correspond to the zeros of this inverse Green's function $G_{\mathrm{ph}}({\bf k},\Omega)$, where we performed an analytic continuation $i \omega_{n} \rightarrow \Omega$. 
\begin{figure*}[t]
  \centerline{\includegraphics{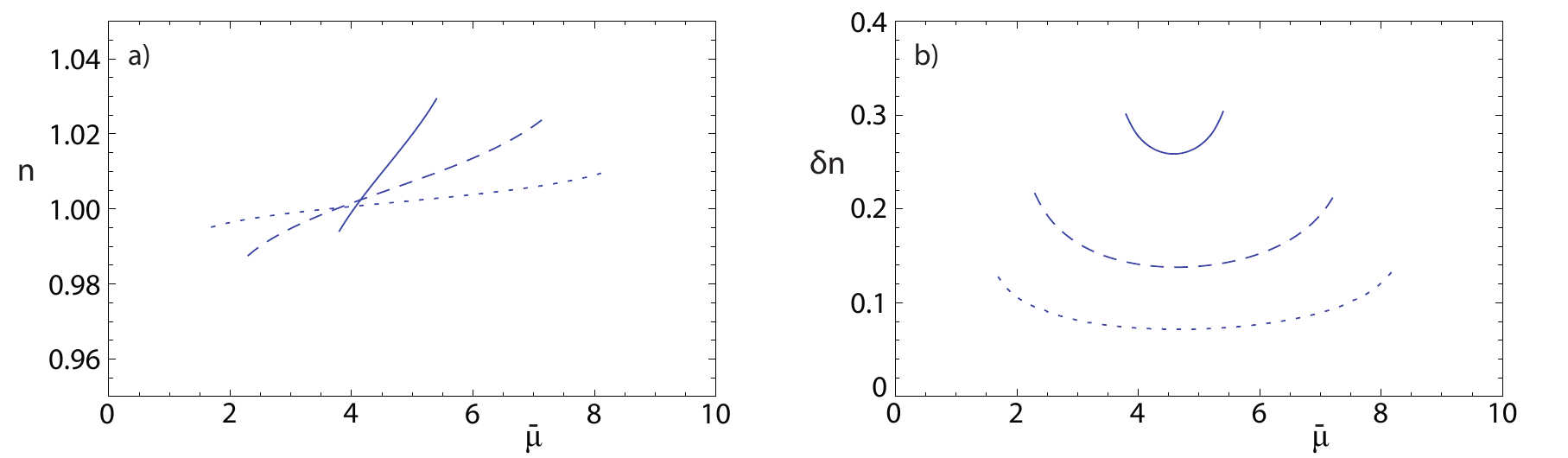}}
 \caption{a) The average number of photons $n$ and b) the corresponding number fluctuation $\delta n$ at zero temperature in the $g=1$ Mott lobe as a function of $\mu / zt$ for ${\bf k} = 0$, $\lambda/\Lambda_{\mathrm{th}} = 1$, $V_{0}/k_{\mathrm{B}}T = 30$, $z t / k_{\mathrm{B}}T \simeq 0.3$ and $U /zt = 11$. The solid, dashed and dotted curve correspond to $\alpha_{\mathrm{lat}} \simeq 9.8 \cdot 10^{-2}$, $\alpha_{\mathrm{lat}} \simeq 4.9 \cdot 10^{-2}$ and $\alpha_{\mathrm{lat}} \simeq 1.6 \cdot 10^{-2}$. Since increasing the value of $\alpha_{\mathrm{lat}}$ also increases the value of $\gamma$, the range of $\bar{\mu}$ where the photons are inside the $g=1$ Mott lobe decreases for increasing $\alpha_{\mathrm{lat}}$. Moreover, the number fluctuation become larger if the value of $\alpha_{\mathrm{lat}}$ increases.}
 \label{fig:NumflucPRA}
\end{figure*}
First, note that for large interactions $U$ the excitations are also at large frequencies. At these large frequencies the self-energy vanishes and we obtain the usual results. However, for intermediate $U$ for which we are inside a Mott lobe and the excitations are still at relatively small frequencies, the self-energy is important. In the following we ingnore the real part of the self-energy, since this part in good approximation only results into a shift of dispersions. 
\newline
\indent By assuming that the excitation are at relatively small energies, we can approximate $\hbar \Sigma^{+}({\bf k},k_{\gamma},\Omega) \simeq - i \alpha \hbar \Omega$. Here $\alpha$ is the small dimensionless damping parameter we calculated in previous work \cite{AW}. Up to linear order in $\alpha$, we obtain
\begin{align}
\hbar\Omega^{\pm}_{{\bf k}} &= (1 - i \alpha_{\mathrm{lat}}) \hbar\omega_{{\bf k}}^{\pm} - \frac{i}{2} \alpha_{\mathrm{lat}} \Bigg{(}U + \mu \\ \nonumber
&\pm \frac{(2 g^{2} - 1) U^{2} - U (\epsilon_{{\bf k}} + \mu (1 + 2 g)) - \epsilon_{{\bf k}} \mu}{\hbar\omega_{{\bf k}}^{+} - \hbar\omega_{{\bf k}}^{-}} \Bigg{)},
\end{align}
where $\alpha_{\mathrm{lat}} =  8 \pi \hbar \alpha /m \omega \lambda^{2}$. Furthermore, $\hbar\omega_{{\bf k}}^{+}$ and $\hbar\omega_{{\bf k}}^{-}$ denote the quasiparticle and quasihole excitations as calculated in Ref.\,\cite{Dries}.
\newline
\indent Moreover, we can also calculate the spectral function that is defined as
\begin{align}
\rho_{\mathrm{ph}}({\bf k},\Omega) = -\frac{1}{\pi \hbar} \mathrm{Im}\left[G_{\mathrm{ph}}({\bf k},\Omega)\right].
\end{align}
In Fig.\,\ref{fig:SpecFM} we show a plot of the spectral function of the photons inside the $g =1$ Mott lobe for ${\bf k} = 0$. The spectral functions has two peaks, one around the quasihole excitations and the other located at the quasiparticle dispersion. Due to the interaction of the photons with the molecules, the peaks are broadened. The width of the peaks is determined by the value of $\alpha_{\mathrm{lat}}$, and the larger this parameter the broader the peaks become. Furthermore, due to the approximation of the self-energy the sum rule is modified and reads \begin{align}
\int d(\hbar\Omega)  \rho_{\mathrm{ph}}({\bf k},\Omega) = \frac{1}{1 + \alpha_{\mathrm{lat}}^{2}},
\end{align}
which is the same sum rule as already encountered in Ref.\,\cite{AW}. Thus for relatively small $\alpha_{\mathrm{lat}}$ the sum rule is in very good approximation satisfied. Note that we can exactly satisfy the sum-rule, by taking the full energy dependence of the selfenergy into account and not using only its low-energy approximation.

\subsection{Number fluctuations}
Apart from the excitations and the spectral function inside the Mott lobes, we can also use the presented theory to calculate the number fluctuations. For a Bose gas in an optical lattice that is described by the Bose-Hubbard model, the true Mott insulator state only exists at zero temperature as at this temperature the number fluctuations inside the Mott lobes vanish. However, for non-zero temperatures thermal fluctuations always induce number fluctuations, and strictly speaking there is no Mott insulator. As we show next, in this system of photons in a dye-filled microcavity even at zero temperature the number of photons in the Mott lobes fluctuates. Therefore, for the photons in the dye-filled optical microcavity the true Mott-insulating state does not exist due to the fluctuations induced by the absorption and emission of photons by the dye molecules.
\newline
\indent To calculate the number of photons inside the Mott lobes, we first determine the thermodynamic potential. After the Hubbard-Stratonovich transformation that decouples the hopping term, the thermodynamic potential consist of two separate parts. The exactly solvable part is given by the eigenvalue of the Hamiltonian $\hat{H}_{0}$ in Eq.\,\eqref{eq:ham} with $\psi = 0$. The other part is described by Eq.\,\eqref{eq:acpsi} and the contribution to the thermodynamic potential can be calculated by integrating out $\psi$ in the Gaussian approximation. Hence,
\begin{align}
\Omega = N_{\mathrm{s}} E_{0} + \frac{1}{\beta} \mathrm{Tr}\left[\log \left(- \hbar \beta G^{-1} \right) \right],
\end{align}
where the inverse Green's function $G^{-1}$ is defined in Eq.\,\eqref{eq:greensplit}, and $E_{0}$ is the energy of the Hamiltonian in the Mott lobe with $g$ photons. Thus,
\begin{align}
E_{0} = \frac{1}{2} U g (g-1) - \mu g. 
\end{align}
Note that the thermodynamic potential also has a contribution from the fact that we change the partition function if we simply add a complete square to the action. 
\begin{figure}[t]
  \centerline{\includegraphics{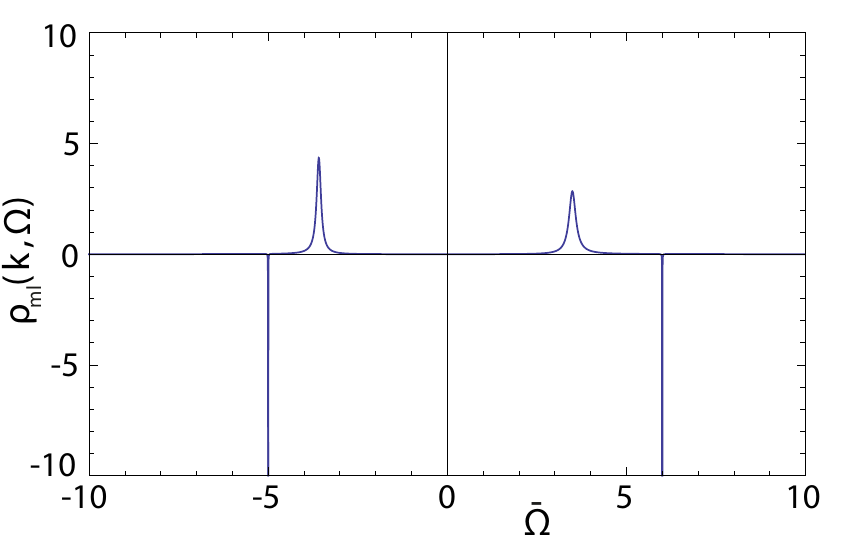}}
 \caption{A plot of the spectral function $\rho_{{\mathrm{ml}}}({\bf k},\Omega)$ inside the $g=1$ Mott lobe as a function of $\bar{\Omega} = \Omega/zt$ for ${\bf k} = 0$, $U /zt = 11$, $\mu / zt = 5$ and $\alpha_{\mathrm{lat}} \simeq 10^{-2}$. The spectral function has a delta-peak contributions at $\Omega = -\mu$ and $\Omega = -\mu + g U$. Furthermore, the non-zero value of $\alpha$ broadens the contributions that are located at the quasihole and quasiparticle excitations.}
 \label{fig:Spectralnumfluc}
\end{figure}
Namely, by performing the Hubbard-Stratonovich transformation we should multiply exactly with one and therefore the thermodynamic potential has an additional contribution that depends on the self-energy induced by the molecules and the hopping parameter. However, both do not have an explicit dependence on the chemical potential of the photons. Therefore, if we calculate densities by taking a partial derivative of the thermodynamic potential with respect to this chemical potential, this additional contribution has no effect. 
\newline
\indent Thus the number of photons per site in the Mott lobe is given by
\begin{align}
n &= -\frac{1}{N_{\mathrm{s}}} \frac{\partial \Omega}{\partial \mu} \\ \nonumber
&= g - \frac{1}{\beta N_{\mathrm{s}}} \sum_{{\bf k},n} G({\bf k},i \omega_{n}) \frac{\partial G^{-1}({\bf k},i \omega_{n}) }{\partial \mu}.
\end{align}
In this expression we cannot simply perform the sum over Matsubara frequencies analytically, since the self-energy has a non-trivial imaginary part. Therefore, we define
\begin{align}
\rho_{\mathrm{ml}}({\bf k},\Omega) = -\frac{1}{\pi} \mathrm{Im}\left[G({\bf k},\Omega^{+}) \frac{\partial G^{-1}({\bf k},\Omega^{+}) }{\partial \mu}\right],
\end{align}
where $\Omega^{+} = \Omega + i \epsilon$ with $\epsilon > 0$  infinitesimally small. In Fig.\,\ref{fig:Spectralnumfluc} we show a typical plot of this spectral function. This function contains four different contributions. There are two delta-peaks at $\Omega = -(\mu - (g-1)U)/\hbar$ and $\Omega = -(\mu - gU)/\hbar$, and there are contributions at the quasiparticle and quasihole excitations that are broadend by the photon-molecule coupling.
\newline
\indent By using the definition of $\rho_{\mathrm{ml}}({\bf k},\Omega)$, we can rewrite
\begin{align}\label{eq:densml}
n = g + \frac{1}{N_{\mathrm{s}}} \sum_{{\bf k}} \int_{-\infty}^{\infty} d(\hbar\Omega) N_{\mathrm{BE}}(\hbar\Omega) \rho_{\mathrm{ml}}({\bf k},\Omega).
\end{align}
Here $N_{\mathrm{BE}}(\hbar\Omega)$ denotes the Bose-Einstein distribution function. Here and in the following we consider the photon gas at zero temperature and only consider quantum fluctuations. This amounts to calculating
\begin{align}
n = g - \frac{\lambda^{2}}{4} \int_{\mathrm{BZ}}\, \frac{d{\bf k}}{(2 \pi)^{2}} \int_{-\infty}^{0} d(\hbar\Omega) \, \rho_{\mathrm{ml}}({\bf k},\Omega),
\end{align}
where we only integrate the momenta over the first Brillouin zone. Recall that we consider a square lattice with spacing $\lambda /2$ and therefore the momenta run from $-2\pi/\lambda$ to $2\pi/\lambda$. Since in the low-energy approximation of the self-energy we obtain ultra-violet divergences, we consider the full expression of the self-energy as obtained in Ref.\,\cite{AW} and numerically evaluate the integrals. However, in the following we still use the parameter $\alpha_{\mathrm{lat}}$ to compare between results for different values of the self-energy. 
\newline
\indent As an example, we now take  $\lambda/\Lambda_{\mathrm{th}} = 1$ and $V_{0}/k_{\mathrm{B}}T = 30$ such that $\beta z t \simeq 0.3$. Furthermore, we calculate the number of photons for different values of the self-energy, which can be obtained by for example changing the density of molecules or changing the detuning. Moreover, these expressions are only valid in the Mott lobes and therefore the values of $\bar{\mu}$ are restricted. In agreement with this theory we use the phase boundaries as calculated by solving $\hbar G^{-1}({\bf 0},0) / z t = 0$, where $\hbar G^{-1}({\bf 0},0) / z t$ is defined in Eq.\,\eqref{eq:Land2}.
\newline
\indent In Fig.\,\ref{fig:NumflucPRA} we show the average number of photons in the $g=1$ Mott lobe at zero temperature for several values of $\alpha_{\mathrm{lat}}$. We observe that inside the Mott lobe the average number of photons is not constant, and therefore the true Mott insulator never exists if the interactions with the dye molecules are included. Furthermore, for increasing values of $\alpha_{\mathrm{lat}}$ there are two effects. First, the range of $\bar{\mu}$ for which the photons are in the $g=1$ Mott lobe decreases as larger values of $\alpha_{\mathrm{lat}}$ also correspond to larger values of $\gamma$. Moreover, the differences between the number of photons for different values of $\bar{\mu}$ inside the plateau become larger, i.e., the slope of the plateau increases. This slope is related to the fluctuations in the average number of photons, namely
\begin{align}
\delta n := \sqrt{\langle \hat{n}^{2} \rangle - \langle \hat{n} \rangle^{2}} =  \sqrt{\frac{\partial n}{\partial (\beta \mu)}} = \sqrt{\frac{\kappa}{z t}},
\end{align}
where we define $\kappa = \partial n / \partial (\beta \bar{\mu})$ as the compressibilty. 
\newline
\indent To obtain the number fluctuations due to quantum fluctuations, we first consider the density as given by Eq.\,\eqref{eq:densml} and we take the derivative with respect to $\beta \bar{\mu}$. Then we only consider quantum fluctuations, by neglecting the Bose-Einstein distribution function of the photons and only integrating over negative frequencies. These number fluctuations are shown in the right figure of Fig.\,\ref{fig:NumflucPRA}. We observe that for increasing values of $\alpha_{\mathrm{lat}}$ the number fluctuations increase. 
\begin{figure}[t]
  \centerline{\includegraphics{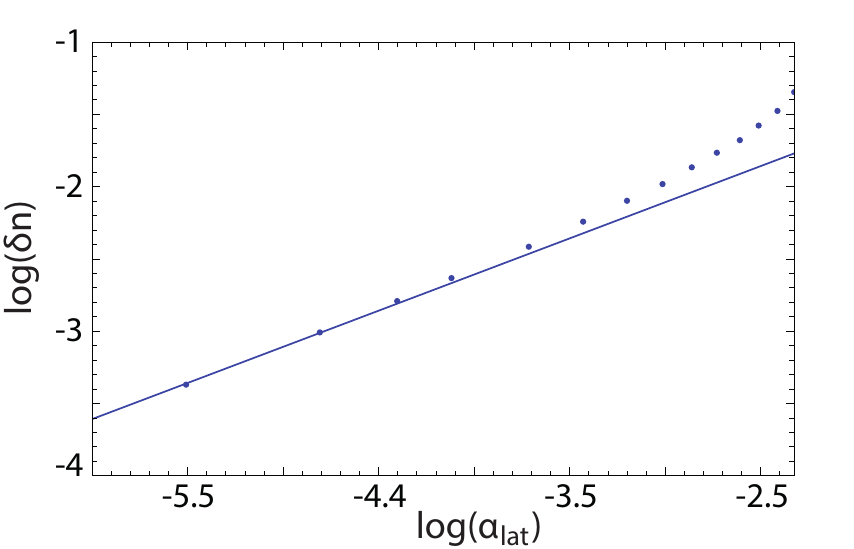}}
 \caption{The minimal value of the number fluctuation $\log(\delta n)$ as a function of $\log(\alpha_{\mathrm{lat}})$ at zero temperature in the $g=1$ Mott lobe for $\lambda/\Lambda_{\mathrm{th}} = 1$, $V_{0}/k_{\mathrm{B}}T = 30$, $z t / k_{\mathrm{B}}T \simeq 0.3$ and $U /zt = 11$. The points represent values of $\bar{\mu}$ for which the minimum in the number fluctuations is determined numerically. The solid line is a fit through the first points of a line with slope $0.5$. For $\alpha_{\mathrm{lat}}$ smaller than roughly $10^{-2}$, we find good agreement with the numerical points.}
 \label{fig:numflucalpha}
\end{figure}
Moreover, for a fixed value of $\alpha_{\mathrm{lat}}$ we obtain that the number fluctuations increase, if we come closer to the phase boundary. Intuitively, this is because the deeper you are in the Mott lobe, on average the fluctuations in the number of photons decreases. 
\newline
\indent To obtain more information about how the value of $\alpha_{\mathrm{lat}}$ affects the number fluctuations, we determine the minimal value of $\delta n$ for different values of $\alpha_{\mathrm{lat}}$. The results of this numerical calculation are shown in Fig.\,\ref{fig:numflucalpha}. We observe that for small values of $\alpha_{\mathrm{lat}}$ in very good approximation $\delta n \propto \sqrt{\alpha_{\mathrm{lat}}}$. If $\alpha_{\mathrm{lat}}$ is larger than roughly $10^{-2}$, this relation is no longer valid and the effect of $\alpha_{\mathrm{lat}}$ on the number fluctuations are larger. We also performed the same calculation for $U / zt = 14$ and we obtained similar results for the scaling of $\delta n$. 
\newline
\indent Although the average number of photons inside a Mott lobe varies, the system is for all practical purposes still a Mott insulator if these number fluctuations are small. To distinguish this region from the regime where the fluctuations are large enough to destroy the Mott insulator, we use our definition of the compressibility. A true Mott insulator is incompressible and this corresponds to $\kappa = 0$ or $\delta n = 0$. However, we anticipate that for $\delta n < 1/2$ it is still possible to make a distinction between different Mott lobes, and therefore we consider this regime practically as the Mott insulator. As can be seen in Fig.\,\ref{fig:NumflucPRA}, this condition is satisfied for  relatively large $\alpha_{\mathrm{lat}}$, even up to values of $10^{-1}$. Thus, in most cases the photons are in good approximation still in the Mott-insulating phase. Moreover, note that these number fluctuations at zero temperature imply that the transition from a superfluid to a Mott insulator is a crossover instead of a quantum phase transition as is the case for cold bosonic atoms in an optical lattice.

\section{Conclusion and outlook}
\label{sec:concl}
In this paper we have investigated the effects of the dye molecules on the superfluid-Mott-insulator phase transition of photons in a dye-filled optical microcavity. First, we derived expressions for the relevant parameters of our theory in terms of the experimental quantities. Hereafter, we considered a simplified model that neglects the fixed longitudional momentum of the photons and only takes into account absorption and emission of photons with zero momentum. We have shown that at the mean-field theory level the effect of the photon-molecule coupling can be captured in a single dimensionless parameter $\gamma$. By performing a self-consistent mean-field theory, we have found that a non-zero expectation value of the annihilation operator of the photons induces coherence between different internal molecular states. Subsequently, we have demonstrated that incorporation of $\gamma$ decreases the size of the Mott lobes. Hereafter, we considered the full model that includes the fixed longitudional momentum and takes into account absorption and emission of photons with non-zero momentum, and we have found generally good agreement between the values of $\gamma$ for this full model and simplified model. However, for small lattice potential depths $V_{0}$ in the full model there is a larger range of lattice spacings where the value $\gamma$ is smaller.
\newline
\indent Moreover, by first integrating out the molecules we calculated both the excitations and the number fluctuations inside the Mott lobes at zero temperature. We obtained that the quasiparticle and quasihole excitation in this system have a finite lifetime, which is visible in the finite width of both contributions in the spectral function. We have demonstrated that the coupling between the photons and dye molecules results into non-zero number fluctuations at zero temperature, and therefore, strictly speaking, the Mott insulator does not exist. However, we have shown that for the most relevant values of $\alpha_{\mathrm{lat}}$, the compressibility is sufficiently small and the system is in good approximation still in the Mott-insulating state. Subsequently, we obtained that for a relatively small coupling the number fluctuations scale with the square root of the dimensionless damping parameter $\alpha_{\mathrm{lat}}$.
\newline
\indent For future research it is interesting to investigate which regions are accessible in experiments. From Ref.\,\cite{PhotLat}, we know that for the lattice potential typical lattice spacings in the micron regime are expected. Since this is of the same order as discussed in this work, we expect that for sufficiently deep lattice potentials the coupling with the molecules is important and could prevent the system from being inside a Mott lobe. For example, it is interesting to compare the state of the photons for different molecular densities, detunings or other dye-specific properties. Furthermore, we have found that in the superfluid phase of the photons the molecules are in a superposition of different internal states. Although our calculation does not incorporate the full rovibrational structure of the molecules, we expect that this phenomenon is also present in the experiment. Therefore, it would be interesting to measure and investigate the behaviour of the molecules if the photons are in the superfluid phase.
\newline
\indent Except for these experimental options, there are also some possibilities for future theoretical work. First of all, in the experimental system there is a harmonic trapping potential and for qualitative agreement this should be taken into account. In a first approximation this can be incorporated in the local-density approximation. Moreover, the presented theory beyond mean-field only takes into account the effect of Gaussian fluctuations at zero temperature. However, in the current experiment the photons are at room temperature. In analogy with the first-order correlation functions and phase fluctuations of a Bose-Einstein condensate of photons under similar conditions, we expect that the thermal fluctuations are also very important \cite{AW2}. Finally, we note that the presented formalism is potentially also useful to describe the effects of relaxation on the superfluid to insulator transition in quantum magnets \cite{QMag}.

\acknowledgments
This work was supported by the Stichting voor Fundamenteel Onderzoek der Materie (FOM), the European Research Council (ERC) and is part of the D-ITP consortium, a program of the Netherlands Organisation for Scientific Research (NWO) that is funded by the Dutch Ministry of Education, Culture and Science (OCW).

\end{document}